\PassOptionsToPackage{numbers,sort&compress}{natbib}
\documentclass[12pt,aps,prd,preprint, onecolumn,nofootinbib,superscriptaddress]{revtex4-2}
\usepackage[left=0.85in,right=0.85in,top=0.9in,bottom=0.9in]{geometry}
\renewcommand{\thesection}{\arabic{section}}

\usepackage{amsmath,amssymb,mathtools}
\allowdisplaybreaks
\usepackage{slashed}
\usepackage{physics}
\usepackage{enumitem}
\usepackage{cancel}
\usepackage{graphicx}
\usepackage{xcolor}
\usepackage{comment}
\usepackage{tikz}
\usetikzlibrary{decorations.pathmorphing}
\usepackage[final,activate={true,nocompatibility},expansion=false]{microtype}
\usepackage[colorlinks=true,linkcolor=blue!70!black,citecolor=blue!70!black,urlcolor=blue!70!black]{hyperref}
\usepackage{url}

\usepackage[most]{tcolorbox}
\newtcolorbox{computation}[1][]{%
    enhanced,
    breakable,
    colback=gray!5,
    colframe=gray!50,
    fonttitle=\bfseries,
    title={Detailed Computation},
    #1
}
\newcommand{\be}[1]{\begin{equation}\label{#1}}
\newcommand{\beq}{\begin{equation}}
\newcommand{\ee}{\end{equation}}
\newcommand{\beqn}[1]{\begin{eqnarray}\label{#1}}
\newcommand{\eeqn}{\end{eqnarray}}

\newcommand{\new}[1]{{\color{black}#1}}
\usepackage{float}

\newcommand{\eeq}{\end{equation}}          

\makeatletter
\AtBeginDocument{%
  \@ifundefined{BibitemShut}%
    {\def\BibitemShut#1{\unskip\csname bibitem#1\endcsname}}%
    {\let\FS@BibitemShut\BibitemShut
     \def\BibitemShut#1{\unskip\FS@BibitemShut{#1}}}%
}
\makeatother
\begin{document}
\title{{\Large \bf Revisiting the $U(1)$-$CP$ tension in QCD} \\
{\Large  from a 1PI-action perspective}}

\author{P. Di Vecchia}
\affiliation{Nordita, KTH Royal Institute of Technology and Stockholm University,
  Hannes Alfv\'ens v\"ag 12, SE-106 91 Stockholm, Sweden}
  \affiliation{The Niels Bohr Institute, Blegadamsvej 17, DK-2100 Copenhagen, Denmark}

\author{F. Sannino}
\affiliation{\mbox{$\hbar$QTC, Quantum Theory Center, \& Danish Institute for Advanced Study (Danish IAS)},
  University of Southern Denmark, Campusvej 55, DK-5230 Odense M, Denmark}
\affiliation{\mbox{Dipt.\ di Fisica ``E.~Pancini'', Universit\`a di Napoli Federico II,
  Via Cintia, 80126 Napoli, Italy}}
\affiliation{Institute for Particle Physics Phenomenology, Durham University,
  South Road, DH1 3LE, Durham, United Kingdom}

\author{G. M.  Shore}
\affiliation{Centre for Quantum Fields and Gravity, Department of Physics,
  Swansea University, Swansea SA2 8PP, United Kingdom}

\author{G. Veneziano}
\affiliation{Theoretical Physics Department, CERN, CH-1211 Geneva 23, Switzerland}
\affiliation{Coll\`ege de France, 11 place Marcelin-Berthelot, F-75005 Paris, France}

\author{S. Yankielowicz}
\affiliation{\mbox{School of Physics and Astronomy, Tel Aviv University,
  Ramat Aviv, 69978 Tel Aviv, Israel}}

\begin{abstract}
We discuss the tension one encounters in trying to solve, simultaneously, the $U(1)$ and  strong-$CP$ problems in QCD with light quarks and a small
$\bar{\theta}$ angle within QCD itself. We do so by considering the low-energy expansion of a
$1$PI effective action $\Gamma$, a functional of a properly chosen set of gauge-invariant (composite)
operators. After enforcing Ward-identity constraints on $\Gamma$, we prove, under minor assumptions, that $CP$ violation at small
non-vanishing $\bar{\theta}$ can be avoided only if the anomaly-induced
contribution to the pseudoscalar mass matrix is completely negligible when compared
to its non-anomalous counterpart---that is, only if the $U(1)$ problem is \emph{not} solved.
The  topological susceptibility in  QCD and a precise generalization of the one in pure Yang-Mills theory emerge as the most important players in establishing the above tension and  in determining the corrections to various known results at leading order in $\bar{\theta}$.

\end{abstract}

\maketitle

\newpage
\setcounter{tocdepth}{2}
\tableofcontents
\bigskip
\newpage

\section{Introduction and outline}\label{sec:outline}

Recently, the question of strong CP violation has been receiving much renewed interest. In view of the fact that present bounds \cite{Abel:2020pzs} on the neutron's electric dipole moment require the QCD $\theta$-angle to be at most $10^{-10}$ \cite{Baluni:1978rf,Crewther:1979pi} the following questions naturally appear:

i) Could there be a non-fine-tuned solution to this problem within QCD itself?

ii) If not, is a Peccei-Quinn symmetry \cite{Peccei:1977hh,Peccei:1977ur} with its accompanying axion \cite{Weinberg:1977ma,Wilczek:1977pj,Bardeen:1977bd} a possible --and perhaps the only possible-- solution?

There have been a few claims \cite{Ai:2020ptm,Nakamura:2021meh,Williams:2026tuw} that the answer to the first question could be affirmative. Objections to such claims have also filled up the archives {(see e.g. \cite{Benabou:2025viy,Aghaie:2026pkf, Khoze:2025auv}). For further references on the current debate  we refer to the review in \cite{Sannino:2026wgx}.}  Without entering into the above debate, we address here the first question from a different perspective. Using 1PI effective action arguments we shall argue that, within QCD  itself\footnote{We are not considering either the coupling of QCD with the electroweak sector or gravity, nor its possible enlargement to include an axion.}, the strong-CP problem at small $\theta$ can only be solved if {\it either} the mass of the lightest quark {\it or} the anomaly contribution to the flavor-singlet pseudo Nambu-Goldstone  boson (PNGB) mass is vanishingly small. The first possibility is excluded by Current-Algebra arguments,  the latter amounts to falling back on the (in)famous $U(1)$ problem~\cite{Weinberg:1975ui}, \cite{Crewther:1979rs}. The tension thus happens to be real, at least within the framework we adopted.

For the sake of rigor, we will discuss this issue by using, rather than a
phenomenological low-energy effective action, the low-energy expansion of
the full $1$PI effective action $\Gamma$ obtained as the Legendre transform of
the generating function $W$ of some chosen set of correlation functions of
gauge-invariant composite operators. This approach was already successfully
applied in the past~\cite{Shore:1991dv,Shore:1990zu,Shore:1991np, Shore:1999tw} in
connection with the so-called proton spin-crisis and with the
$\eta' \to 2\gamma$ decay.

We will also apply the above method  to determine the constraints
to be satisfied by $\Gamma$ as a consequence of exact, softly broken, and
anomalous Ward identities. This will allow us to discuss the question of
whether some commonly used effective actions do or do not satisfy these
constraints.

The rest of the paper is organized as follows:
In Section~\ref{sec:general} we will prove a useful (and often implicitly used)  lemma on $\Gamma$
before renormalization, and work out, at the
renormalized level, its implications together with those coming from
(anomalous and non-anomalous) Ward identities.

In Section~\ref{sec: theorem} we apply this formalism to give a  general ``proof" of the $U(1)-CP$-tension ``theorem" alluded to at the beginning \new{recovering, en-passant and beyond the large-$N$ limit, the known connection \cite{Crewther:1979pi,DiVecchia:1980yfw,Witten:1980sp} between} the coefficient of the $CP$-violating Lagrangian and the topological susceptibility in full (unquenched) QCD.

In Section~\ref{sec:thetadep}  we will present some amusing relations on the $\theta$-dependence of various observables.

In Section~\ref{sec:examples} we will see how the above general results apply to a few well-known examples: i) the effective Lagrangian of  the conventional 't Hooft large-$N_c$ ($N$ hereafter)  limit, which has been widely discussed in the literature \cite{Witten:1979vv,Veneziano:1979ec,DiVecchia:1980yfw,Witten:1980sp,Rosenzweig:1979ay,Nath:1979ik,Kawarabayashi:1980dp}; ii) the original 't Hooft effective Lagrangian \cite{tHooft:1976snw} and some simple generalizations thereof; and iii) its analog in the orientifold large-$N_c$ limit \cite{Armoni:2003gp}. In each case we will work out their consequences for the $U(1)-CP$-tension.

In Section~\ref{sec:discussion} we will summarize our main results. To help the reader, a couple of technical points are shifted to two short appendices.

\section{General properties of $1$PI effective actions}
\label{sec:general}
 As  mentioned in the introduction, our approach is based on the concept of $1$PI effective actions $\Gamma (O_i)$ as functionals of a set of composite gauge invariant operators $O_i$. We use the plural (``effective actions") to emphasize that there are many possible choices for the arguments $O_i$ of $\Gamma$. However, also the rules of how to deal with $\Gamma$ depend on that choice, and precisely in such a way that the connected correlators of (a common set of) the composite fields do not depend on that choice\footnote{The point is that the connected correlators are obtained from $\Gamma$ by simply considering ``tree" diagrams involving only the propagators of the chosen set of composite fields. Poles due to propagators of other composite fields are already included in $\Gamma$ and can spoil its analytic properties if possibly light degrees of freedom are not part of the arguments of $\Gamma$. We refer to ~\cite{Shore:1991dv,Shore:1990zu,Shore:1991np,
 Shore:1999tw} for further details on this formalism which also includes the possibility, not exploited in this paper, to have  mixed (sometimes referred to as ``Zumino") effective actions in which only a subset of the variables undergoes a Legendre transformation.}

\subsection{A lemma on effective actions for composite operators}
\label{sec:lemma}
Consider a generic gauge theory (in particular QCD)\new{, whose action $S$ depends on a set of parameters $\lambda_i$,} and introduce the
generating function of the connected correlators of a set of gauge-invariant
composite operators $O_i$:
\begin{equation}\label{eq:W-def}
\exp\!\bigl(iW(J_i\new{;\lambda_i})\bigr)
= \int d(\ldots)\;\exp\!\left(
  iS(\ldots\new{;\lambda_i}) + i\sum_i O_i\, J_i
\right),
\end{equation}
where the dots represent the fundamental fields of the gauge theory
(including ghosts) and we have adopted the convention that $\sum_i$
includes integration over spacetime.

Let us then define the 1PI effective action as a functional of the
``classical'' fields $O_i^{cl.}$  by the usual Legendre transform:
\begin{equation}\label{eq:Legendre}
\Gamma(O_i^{cl.}\new{;\lambda_i}) = W(J_i\new{;\lambda_i}) - \sum_i J_i\,O_i^{cl.}\;,
\qquad
O_i^{cl.} \equiv \frac{\delta W}{\delta J_i}
\;\;\Rightarrow\;\;
J_i = J_i(O_i^{cl.})\;,
\qquad
\frac{\delta\Gamma}{\delta O_i^{cl.}} = -J_i\,.
\end{equation}
As a result we can also write
\begin{equation}
\label{eq:Gamma}
\exp(i \Gamma(O_i^{cl.}\new{;\lambda_i})) = \int d(\dots) \exp\left(i S(\dots\new{;\lambda_i}) + i \sum_i  \frac{\delta \Gamma}{\delta O_i^{cl.}}(O_i^{cl.} -O_i ) \right)
\end{equation}

Consider now the case in which the fundamental action $S$ depends
\emph{linearly} on some parameters $\lambda_i$ multiplying the composite
fields $O_i$. We claim that, as a result, also $\Gamma$ depends linearly
on those same parameters, i.e.
\begin{equation}
\label{eq:linearity-lemma}
S(\dots, O_i;\lambda_i + \delta_i)
= S(\dots, O_i;\lambda_i) + \sum_i \delta_i\,O_i
\;\;\Longrightarrow\;\;
\Gamma(O_i^{cl.}\new{;}\lambda_i+\delta_i)
= \Gamma( O_i^{cl.}\new{;}\lambda_i) + \sum_i \delta_i\,O_i^{cl.}\,.
\end{equation}
 \new{The proof is immediate. Since the action is, by assumption, linear in the $\lambda_i$, it follows directly from (\ref{eq:W-def}) that
\begin{equation}\label{eq:Wshift}
W(J_i;\lambda_i+\delta_i) = W(J_i+\delta_i;\lambda_i)\,,
\end{equation}
hence $O_i^{cl.}(J;\lambda+\delta) = O_i^{cl.}(J+\delta;\lambda)$: the two theories have the same classical fields $O_i^{cl.}$ when their sources differ by $\delta_i$. Denoting by $J_i$ the sources that produce the given $O_i^{cl.}$ at parameters $\lambda_i$, the Legendre transform (\ref{eq:Legendre}) then gives
\begin{equation}\label{eq:lemmaProof}
\Gamma(O_i^{cl.};\lambda_i+\delta_i) = W(J_i;\lambda_i) - \sum_i (J_i-\delta_i)\,O_i^{cl.}
= \Gamma(O_i^{cl.};\lambda_i) + \sum_i \delta_i\,O_i^{cl.}\,,
\end{equation}
which is the claim. Note that the lemma holds for a parameter $\lambda_i$ only if the operator it multiplies in the (regularized) action is among the Legendre-transformed arguments of $\Gamma$, with the corresponding source coupled to exactly that operator.}

This result applies, in particular, to QCD, where the  $\lambda_i$
can be taken to consist of:
\begin{itemize}
\item the (complex) quark mass matrix $m_{ij}$, multiplying the scalar and pseudoscalar quark bilinears   $ {\cal M}_{ij} \equiv \bar{q}_{R,i}q_{L,j} $ (with $i, j = 1, 2 \dots N_f$) and its hermitian conjugate,
\item the $\theta$ angle, multiplying the topological charge density
  $Q \equiv \frac{1}{32\pi^2}G\tilde{G}$.
\end{itemize}
\new{As discussed in the next subsection, the linearity in $m$ and $\theta$ survives renormalization.}

For the purposes of this paper it will suffice to consider $\Gamma$ as a functional of only ${\cal M}^{cl.}$ and $Q^{cl.}$ At the unrenormalized level we get:
 \beq
\label{eq:GammaUQ}
 \Gamma({\cal M}^{cl.}, Q^{cl.}; m, \theta) =  \Gamma({\cal M}^{cl.}, Q^{cl.}; 0, 0) - Tr (  m {\cal M}^{cl.}+ h.c.) + \theta Q^{cl.}
\eeq
Let us stress that this extremely simple result for the $m$ and $\theta$ dependence of $\Gamma$ only holds if we associate the classical fields ${\cal M}^{cl.}$ and $Q^{cl.}$ with the original quantum fields ${\cal M}$ and $Q$. In particular, ${\cal M}$ corresponds to a set of $2 N_f^2$ real fields half of which are scalars and half pseudoscalars.
Therefore, for the time being, we are in the so-called linear realization of the flavor symmetry.
From here on the superscript $cl.$ will be omitted whenever this cannot generate
confusion.

\subsection{Renormalization and anomaly constraints for QCD}
Let us now interpret~\eqref{eq:GammaUQ} in terms of (finite)
renormalized fields and parameters.
\begin{itemize}
\item \textbf{$\theta$ and $Q$:} The situation is simple since the topological charge and $\theta$ are RG-invariant, while the topological charge density gets mixed with the divergence of the $U(1)$ axial current in a very simple way \cite{Shore:1991np}.
\item \textbf{$m$ and ${\cal M}$:} We know that their product is
  renormalization-group invariant (RGI), so we can rewrite the same term
  in the effective action in terms of the renormalized mass matrix and quark
  bilinears.
\end{itemize}
\new{More precisely, we employ dimensional regularization with mass-independent $\overline{\rm MS}$ subtraction and the BMHV prescription for $\gamma_5$ \cite{tHooft:1972tcz,Breitenlohner:1977hr}, supplemented by the finite renormalizations of the axial currents and of the pseudoscalar densities that restore the non-anomalous chiral Ward identities and the conventional form of the anomalous singlet one \cite{Adler:1969er,Larin:1993tq}. In such a scheme $m {\cal M}$ is RG-invariant, so that the linearity in $m$ holds for the renormalized mass. The renormalization of the topological charge density has been established to all orders in perturbation theory \cite{Espriu:1982bw,Luscher:2021bog}: no divergent multiplicative renormalization is required, the only effect being an additive mixing with the divergence of the flavor-singlet axial current. For constant $\theta$ the renormalization constants are independent of $\theta$ and, at non-zero quark masses, the space-time integral of the divergence of the axial current vanishes, there being no massless pole in the singlet channel. The mixing therefore does not affect the zero-momentum quantities used below, and the linearity of $\Gamma$ in $\theta$, which follows from the Legendre-transform construction rather than from the non-renormalization of the anomaly, survives renormalization. The same properties hold non-perturbatively in any $CP$-conserving regularization in which $\int Q$ is integer-valued and all phases of the mass matrix can be rotated exactly into $\theta$, e.g.\ a lattice regularization with Ginsparg--Wilson fermions \cite{Ginsparg:1981bj,Hasenfratz:1998ri,Luscher:1998pqa}.}

 At the same time we should impose that $\Gamma$ satisfies, at its own tree level, the anomalous as well as the non-anomalous Ward identities that follow from the symmetries of the classical action and from their breaking due to quantum effects. These WI's take a particularly simple form when the fields appearing in the WI's are the same as the arguments of $\Gamma$.

 We will do this for a $\Gamma({\cal M},Q)$ incorporating the axial anomaly and the $\theta$ angle. In order to address the question of the trace anomaly it would be possible to start from an action $\Gamma({\cal M},Q,F)$ that also includes the composite field \new{$F \equiv G^2$}. Since this will not be needed for the purposes of this paper we will just use a $\Gamma({\cal M},Q)$ and refer the reader to the literature \cite{Schechter:1980ak,Migdal:1982jp} for a discussion of how to impose  the trace anomaly and renormalization group constraints.

 The $U(1)$ anomalous WI implies that a $U(1)$ rotation $2\beta$ of  ${\cal M}$ must be equivalent to a shift $2 N_f \beta$ in $\theta$. We deduce that linearity in $\theta$ implies\footnote{It might naively appear that (\ref{eq:GammaUQan}) necessarily gives the large-$N$ form of the anomaly contribution to $\Gamma$. We will see in Sect. (\ref{sec:examples}) that this is not the case.}:
 \beq
\label{eq:GammaUQan}
 \Gamma({\cal M}, Q; m, \theta) =  \Gamma_{inv.}({\cal M}, Q; 0, 0) - Tr (  m {\cal M} + h.c.) + (\theta - \frac{i}{2} ~tr \log ({\cal M}/{\cal M}^{\dagger}) ) Q
\eeq
where $ \Gamma_{inv.}$ is invariant under the full $U(N_f)_R\otimes U(N_f)_L$ global symmetry group.

In this paper we are mainly concerned with the pseudoscalar spectrum of the theory and with CP-violation due to a small non-vanishing $\theta$ angle.
Our entire discussion will thus be based on the low-energy limit of the effective action $\Gamma({\cal M},Q)$ as given in (\ref{eq:GammaUQan}). Using non-anomalous field redefinitions we can also bring the mass matrix $m$ to real diagonal form \footnote{We are also assuming $m_i >0$ so that $\bar{\theta}$ will be close to 0 rather than to $\pi$.}, $m_{ij} = \delta_{ij} m_i$, at the cost of replacing $\theta$ with $\bar{\theta} = \theta - \arg \det m$. Hereafter, $\theta$ will actually refer to $\bar {\theta}$.

\section{A ``Theorem" for the general case}
\label{sec: theorem}
Let us now consider the general QCD effective action $\Gamma({\cal M},Q)$  introduced in Sect. (\ref{sec:general}) with different useful splittings of its various pieces:
\begin{eqnarray}
\label{eq:GammaUQan-gen}
 && \Gamma({\cal M}, Q; m, \theta) =\Gamma_{inv}+ \Gamma_m + \Gamma_{an} \nonumber \\
 && \equiv  \Gamma_{inv}({\cal M}, Q; 0, 0) - Tr (m ({\cal M} + {\cal M}^{\dagger}) )+ \left(\theta - \frac{i}{2} ~tr \log ({\cal M}/{\cal M}^{\dagger}) \right) Q ; \nonumber \\
 &&\Gamma_{inv}({\cal M}, Q; 0, 0) = \Gamma_{kin}({\cal M}, Q; 0, 0) - V_{inv}({\cal M}, Q; 0, 0) \nonumber \\
 && V({\cal M}, Q) \equiv  V_{inv}({\cal M}, Q) +Tr (m ({\cal M} + {\cal M}^{\dagger} ))-  \left(\theta - \frac{i}{2} ~tr \log ({\cal M}/{\cal M}^{\dagger}) \right) Q
 \, ,
\end{eqnarray}
where we have further split $\Gamma_{inv}$ into a ``kinetic" (i.e. derivative) and a potential (i.e. zero momentum) term and we have also defined the total potential $V$. We are interested in the low-energy, small-mass and small-$\theta$ limit and have assumed $\Gamma_{inv}({\cal M}, Q; 0, 0)$ to be regular around vanishing momenta\footnote{Actually there can be some non-analyticity in $\Gamma$ from loops of exactly massless NG bosons, but this is, by assumption, not our case, since it would make the strong-$CP$ problem disappear.  We also assume the ratio $N_f/N$ to be small enough for us to be below not only the asymptotic-freedom limit, but also sufficiently far below its lower edge \cite{Banks:1981nn,Sannino:2004qp,Dietrich:2006cm,DeGrand:2015zxa}. The location of the window depends on the fermion representation, and is given for a general representation in \cite{Dietrich:2006cm}. Near that edge the approximate scale invariance of the dynamics can leave a parametrically light scalar singlet, a dilaton, in addition to the pseudo-Nambu-Goldstone spectrum \cite{Bando:1986bg,Dietrich:2005jn,Appelquist:2010gy,Antipin:2011aa}. Were such a state present, the low-energy description would have to be enlarged to accommodate it \cite{Grinstein:2011dq,Hansen:2016fri,Golterman:2024vra}.}.

We further assume $V_{inv}({\cal M}, Q; 0, 0)$ to be analytic around $Q=0$, so that it can be expanded in even powers of $Q$ around $Q=0$:
\beq
\label{eq:Gammainvesp}
   V_{inv}({\cal M},Q; 0, 0) = V_{inv}({\cal M},0) + \frac{Q^2}{2}  V_{QQ} ({\cal M}, 0) + {\cal O}(Q^4)
\eeq
This should be a safe assumption for the case $\theta \ll 1$ under discussion, since around $\theta =0$ the ground state should be unique, CP conserving and gapped~\cite{Vafa:1984xg}: possible branch changes are only expected at $\theta \sim \pi$.

According to the rules of 1PI effective actions we can read off physical observables connected with a particular field appearing in $\Gamma$ only after integrating out all the others through their classical equations. If we are interested in $q\bar{q}$ correlators it will be necessary therefore to eliminate $Q$ through its classical equations {\it before} using those for ${\cal M}$.
Denoting by $\hat{Q}({\cal M})$ such a solution at zero momentum  we immediately find:
\beq
 \label{hatQ}
\left(\theta - \frac{i}{2} ~tr \log ({\cal M}/{\cal M}^{\dagger}) \right) = \frac{\partial V_{inv}}{\partial Q}({\cal M}, \hat{Q}) \equiv V_{inv,Q}({\cal M}, \hat{Q})
\eeq
from which we can in principle get both $\hat{Q}$ and $\Gamma({\cal M}) \equiv \Gamma({\cal M},\hat{Q}({\cal M}))$ in terms of ${\cal M}$ (and the parameters $m_i, \theta$).

We should contrast $\hat{Q}({\cal M})$ with the set $\langle Q \rangle, \langle {\cal M} \rangle$ which is defined, instead, as
the solution of the full set of vacuum equations.
In order to write those,  it is useful to parametrize the matrix ${\cal M}$ as the product of a Hermitian and a unitary matrix:
\beq
\label{parM}
{\cal M} = H U ~;~ H = H^{\dagger}~;~ U U^{\dagger} = 1 ~;~ U = e^{i \phi}
\eeq
where we have further parametrized $U$ in terms of the Hermitian matrix $\phi$.  With  the vacuum expectation value of $H$ corresponding to half of the quark condensate. The latter can be further decomposed as:
\beq
\label{parPhi}
\phi_{ij} = \sum_{a=1}^{N_f^2} \lambda^a_{ij} \phi_a = \delta_{ij} \phi_i + \sum_{a \ne i} \lambda^a_{ij} \phi_a
\eeq
namely by separating, in a standard way,  the phases $\phi_i$ belonging to the Cartan subalgebra of $U(N_f)$ from those which do not.

The advantages of this parametrization are  that $V_{inv}$ cannot depend on the $\phi_{ij}$, but  only on $H$, while the anomaly term depends only on
$tr \log ({\cal M}/{\cal M}^{\dagger}) = 2 i \sum_i \phi_i = 2i Tr \phi$. Consequently Eq. (\ref{hatQ}) for $\hat{Q}$ simplifies as:
\beq
\label{hatQs}
\left(\theta + Tr \phi \right) =  V_{inv,Q}(H, \hat{Q}) \quad \Rightarrow  \quad \hat{Q} = \hat{Q}(H, \theta + Tr \phi ) \, .
\eeq

We now have to set to zero the variation w.r.t. ${\cal M}_{ij}$ hence w.r.t. the matrix elements of the two Hermitian matrices $H$ and $\phi$.
Our  first claim is that, for a real diagonal mass matrix $m$, there is a consistent stationary point at which also ${\cal M}_{ij}$ is diagonal. The proof is given in Appendix A. Yet this does {\it not} imply that there are no lower-lying minima where some off-diagonal entry of ${\cal M}$ gets a VEV. Our assumption to stick to a diagonal $\langle {\cal M} \rangle$ amounts to the (standard) one that the vector symmetry $(U_V(1))^{N_f}$, which is preserved by the mass term, is {\it not} spontaneously broken.

 {Now, varying $V$ w.r.t. $\phi_{i}$, we find the following system of equations for the vev's $\langle \phi_i \rangle$ and $\langle Q \rangle$ of $\phi_i$ and $Q$:}
\begin{eqnarray}
\label{solangles}
  &  2 H_i m_i \sin \langle \phi_i \rangle + \langle Q \rangle =0 ~;~ {i = 1, 2 \dots N_f} \\
  & \left(\theta + \sum  \langle \phi_i \rangle  \right)  = V_{inv,Q}(H_i, \langle Q \rangle )
\end{eqnarray}
 where $H_i$ are still to be determined through their own stationarity conditions.

The final step is to show that $H_i$ is proportional to the unit matrix up to corrections ${\cal O}(\frac{m_i}{\Lambda_{QCD}})$ and that these corrections do not affect the leading order results in a small-$m_i$ expansion\footnote{A similar statement should be made about the role of heavy quarks ($m \gg \Lambda_{QCD}$). In that case higher order corrections will ensure their decoupling from the low-energy dynamics (Appelquist-Carazzone theorem \cite{Appelquist:1974tg}).}. The proof of this statement is also relegated to Appendix A.
The bottom line of this discussion is that, for the quantities  of interest to us, we may take $H$ to be frozen at its expectation value $\langle H_{ij} \rangle = \delta_{ij} H$ of the massless theory, which amounts to saying that we can move to the (more commonly used) non-linear realization of spontaneously broken chiral symmetry.

The above-mentioned assumptions relevant to prove the main claim of this paper can be summarized as follows:
   (i) $V_{inv}$  is analytic in $Q$ around $Q=0$ \new{(this does not exclude a multi-branched $\theta$-dependence: the large-$N$ potential of Sect.~\ref{sec:examples} is multi-branched in $\theta$ yet analytic at $Q=0$, cf.\ \cite{Witten:1980sp,Gaiotto:2017yup})}; {(ii) There is a mass gap above the PNGBs masses except, possibly, for the pseudoscalar singlet.}  
   \bigskip

{\bf \noindent Claim: \\
In massive QCD CP-violation at small non-vanishing $\bar{\theta}$ can only be avoided if the anomaly-induced PNGB mass matrix is negligible wrt its non-anomalous counterpart, i.e. if the U(1) problem is {\it not} solved.}
\bigskip

In order to prove this statement let us compute, separately, the low-lying spectrum and CP-violation that follow from the above action. 
We first factor out from $U$ its VEV $\langle U \rangle$ and write:
\beq \label{Ufact}
U = \langle U \rangle \times e^{i \Phi} \equiv {\rm diag.}(e^{i \langle \phi_i \rangle} ) \times e^{i \Phi}
\eeq
so that $\Phi$ represents the PNGB fluctuations.  

Let us start with the pseudoscalar mass spectrum. As already mentioned, in order to do that we have to eliminate $Q$ by replacing it with $\hat{Q}(U)$ from inverting (\ref{hatQs}).
It is convenient to split the full potential as follows:
\beq
\label{Vsplit}
V(U) = V_m(U) + W(U); \qquad   W(U) \equiv  V_{inv}(\hat{Q}) -  (\theta + Tr \phi) \hat{Q}
\eeq
With the parametrization (\ref{Ufact}) the mass term can be rewritten as
\beq
\label{massterm}
V_m = H Tr (m   e^{i \langle \phi \rangle} ~ e^{i \Phi} + h.c) =  2 H Tr \left[m (\cos \langle \phi \rangle  \cos \Phi - \sin \langle \phi \rangle \sin \Phi ) \right]
\eeq
where we have used the cyclic invariance of the trace and the fact that $m$ and $\langle U \rangle$, being diagonal, commute.

Concerning the $W$ piece of the potential we note that it is just a function of $(\theta + Tr \phi)$ and that its derivative with respect to $Tr \phi$ (or $\theta$) simplifies thanks to (\ref{hatQs}):
\beq
\label{Wder}
 \frac{\partial W}{\partial (Tr \phi)} = - \hat{Q} - (\theta + Tr \phi) \frac{\partial \hat{Q}}{\partial (Tr \phi)} + \frac{\partial V_{inv}}{\partial \hat{Q}} \frac{\partial \hat{Q}}{\partial (Tr \phi)} = - \hat{Q}
\eeq

In order to compute the spectrum we need the quadratic part of $V$ (as well as the kinetic terms). For the PNGB's outside the Cartan sub-algebra the only contribution comes from the mass term and one recovers the famous GMOR relation\footnote{Note that to get the correct sign for the PNGB's square masses one must have $H = - |H| < 0$.} with ${\cal{M}} = {\bar{q}}_{R,i} {q}_{L,i} $  up to the replacement of $m$ by $m \cos \langle \phi \rangle$. If this approximation also applied to the PNGB in the Cartan subalgebra we would simply rediscover the $U(1)$ problem.  

The only way to solve it is to appeal to the contribution coming from $W$.  {The latter is an even function of its argument  which  is $  \theta + Tr \langle \phi \rangle + Tr \Phi$. }  {We now crucially use the fact that, at small $\theta$ the combination $ \theta + Tr \langle \phi \rangle $ is also small (actually it is smaller than $\theta$ by a factor $m/\Lambda_{QCD}$, see (\ref{thetaSigma}) below). That means that we can expand $W$ as a Taylor expansion in $ \theta + Tr \langle \phi \rangle$ at finite $Tr \Phi$ and keep just the leading term. The situation is quite different if we look at even or odd powers of $ Tr \Phi$. } {For the singlet mass term, and for other CP-even terms, we are interested in extracting even powers of $Tr \Phi$. In that case the leading term has no $(\theta + Tr \langle \phi \rangle)$-dependence. Using (\ref{Wder}), we can express its $Tr \Phi$-expansion in terms of the Taylor expansion of $\hat{Q}$:
\beq
\label{Weven}
W =  W(Tr \Phi) + {\cal O}((\theta + Tr \langle \phi \rangle)^2) = - \sum_{n=0} \frac{\hat{Q}^{(2n+1)}(0)}{(2n+2)!}(Tr \Phi)^{2n+2} + {\cal O}((\theta + Tr \langle \phi \rangle)^2)
\eeq}
 {In particular the quadratic term in $Tr \Phi $ we are after is related to $\hat{Q}$ by:
\beq
\label{Wder2}
\frac{\partial^2 W}{\partial (Tr \Phi)^2}|_{Tr \Phi=0} = - \frac{\partial \hat{Q}}{\partial \theta}|_{Tr \Phi=0} = - \frac{1}{V_{QQ}(0)} \equiv A
\eeq
so that the quadratic term in $Tr \Phi $ is well approximated by:
 \beq
 \label{Wapp}
W \sim \frac{A}{2} (Tr \Phi)^2 \quad {\rm with} \quad Tr \Phi = \frac{\sqrt{2N_f}}{F_{\eta_0}}\,\eta_0 \quad  \Rightarrow \quad m_{\eta_0}^ 2 = \frac{2 N_f A}{F_{\eta_0}^2}  \ , 
\eeq
where, in the last step, we have taken the massless limit and have introduced the $\eta_0$  decay constant which is in general different from $F_{\pi}$ as a result of a possible  double-trace kinetic term\footnote{We also recall that, unless one takes quark masses to be all equal, the mass eigenstates in the Cartan subalgebra are mixed combinations of pure $SU(N_f)$ representations. In particular, the physical $\eta'$ should not be confused with the pure flavor singlet $\eta_0$.} (see \cite{Shore:1991dv} for a thorough discussion of its RG properties). } \new{Since $A$ is defined at zero momentum while the mass is the position of the pole, (\ref{Wapp}) holds up to corrections of relative order $m_{\eta_0}^2/\Lambda_{QCD}^2$ from the momentum dependence of the 1PI two-point function of $Q$. These corrections vanish precisely in the regime $A \ll m |H|$ in which $CP$ violation would be suppressed, so that our conclusions are unaffected.} 

After inserting the properly normalized fields through the various decay constants $f_{ij}$ the additional contribution (\ref{Wapp}) to the  mass matrix can solve the $U(1)$ problem for a suitable negative value of $V_{QQ}$ very much in the spirit of the large-$N$ solution where it is known \cite{Witten:1979vv} that the quadratic term in $Q$ must have the ``wrong'' sign.

\noindent
{Similarly, for the $(Tr \Phi)^4$ term we have:
\beq
 \label{W4}
W \sim -\frac{\hat{Q}^{(3)}(0)}{4!}(Tr \Phi)^4 
\eeq
contributing to the low energy limit of $\eta_0 \eta_0$ elastic scattering.}

In conclusion, eq. (\ref{Weven}) generalizes, away from the large-$N$ limit, the WV relation \cite{Witten:1979vv,Veneziano:1979ec} with $A$ replacing the Yang-Mills topological susceptibility.

Let us now turn to the computation of the CP-odd part of $V$. CP-odd terms originate from all the terms in $\Gamma(U,Q)$, even from those that are CP-even, since we have to take into account the non-trivial condensate phase alignment at non-zero $\theta$.

Consider first the (CP-conserving) mass term. In terms of the shifted fields $ {\Phi}$ it generates a CP-odd term:
\beq
\label{CPmass}
 - 2 H  Tr (m \sin \langle \phi \rangle \sin \Phi ) =   \langle Q \rangle  \times Tr (\sin \Phi )
\eeq
where we have used (\ref{solangles}).

 {Consider next the contributions of $W$ as given by (\ref{Wder}).
This time the leading term in the small-$(\theta + Tr \langle \phi \rangle)$ is linear in $(\theta + Tr \langle \phi \rangle)$ and we obtain:
\begin{eqnarray}
\label{Wodd}
W  &=&  (\theta + Tr \langle \phi \rangle) ~ W^{(1)}(Tr \Phi) + {\cal O}((\theta + Tr \langle \phi \rangle)^3)  = - (\theta + Tr \langle \phi \rangle) ~ \hat{Q}(Tr \Phi) + {\cal O}((\theta + Tr \langle \phi \rangle)^3) \nonumber \\ &=& - (\theta + Tr \langle \phi \rangle) ~\sum_{n=0} \frac{\hat{Q}^{(2n+1)}(0)}{(2n+1)!}(Tr \Phi)^{2n+1} + {\cal O}((\theta + Tr \langle \phi \rangle)^3)
\end{eqnarray}}

The linear term in $Tr \Phi$ is simply given by $\hat{Q}$ evaluated at $Tr \Phi =0$, i.e.\ by $\langle Q \rangle$.  Such a linear term nicely combines with (\ref{CPmass}) to give $\langle Q \rangle   \times Tr [\sin \Phi - \Phi] $ neatly cancelling the linear term in the CP-odd potential, as should be the case on the solution to the field equations.

\noindent
{As indicated in (\ref{Wodd}), a whole tower of higher-order CP-odd terms is generated by $\hat{Q}$.  Unlike (\ref{CPmass}), (\ref{Wodd}) depends on the  arbitrary form of $V_{inv}$. Nevertheless, this arbitrariness only affects vertices that involve just $\eta_0$. We will provide examples of this model dependence in Sect. (\ref{sec:examples}).}

We have thus seen how $\hat{Q}$ and $\langle Q \rangle $ are the main players in determining both the size of CP-violation and the possible resolution of the $U(1)$ problem. What is the meaning of these two quantities? How are they related? 

$\langle Q \rangle $  controls the strength of CP-violation and, for small $\theta$, is related to the topological susceptibility in QCD through:
\beq
\label{Qchia}
\langle Q \rangle  = - \chi_{QCD} \theta + {\cal O}(\theta^3)  \ ,
\eeq
which leads us to define
 \beq
\label{Qchi}
  ~ \chi_{QCD}  = - \frac{\partial \langle  Q  \rangle }{\partial   \theta  }|_{\theta =0}  = \new{+} \frac {\partial^2 V}{(\partial \theta)^2}|_{\theta =0} = - \int d^4 x \langle Q(x) Q(0) \rangle|_{\theta =0} \ .
\eeq
On the other hand the singlet mass is controlled by $V_{QQ}$ which looks, a priori, quite unrelated to $\chi_{QCD}$.
However, using (\ref{hatQ}) at small $\theta$, we see that:
\beq
\chi_{QCD}~ \theta =  A  \,  (\theta + Tr \langle \phi \rangle)
\eeq

 The quantity $(\theta + Tr \langle \phi \rangle)$ can be estimated at
small $\theta$. Linearizing Eq.~(\ref{solangles}) ($\sin \langle \phi_i
\rangle \to \langle \phi_i \rangle$, $H = -|H|$) gives $\langle \phi_i
\rangle = \langle Q \rangle / (2 |H| m_i)$, while Eq.~(\ref{hatQs}) at the
vacuum, with $V_{inv,Q} = V_{QQ}\, Q$ and $A \equiv - 1/V_{QQ}(0)$, gives
$\langle Q \rangle = - A \left(\theta + Tr \langle \phi \rangle\right)$.
Summing the first relation over $i$ and inserting the second,
\beq
Tr \langle \phi \rangle = \langle Q \rangle \sum_i (2|H| m_i)^{-1}
= - A \left(\theta + Tr \langle \phi \rangle \right) \sum_i (2|H| m_i)^{-1} \, ,
\eeq
and solving for $(\theta + Tr \langle \phi \rangle)$ one finds the result:

\beq\label{thetaSigma}
(\theta + Tr \langle \phi \rangle) = \theta \times \frac {1}{ 1 + A \sum (2 |H |m_i)^{-1} } + {\cal O}(\theta^3)
\eeq
Combining (\ref{Wapp}), (\ref{CPmass}), (\ref{Wodd}) we conclude that:
\beq
V_{CP-odd} \sim \theta \times \chi_{QCD}  = \theta \times \frac {A}{ 1 + A \sum (2 |H |m_i)^{-1}}
\eeq
Therefore,  since $A$ must be larger than $|H |m \sim - m \langle \bar{q}_i q_i \rangle$  in order for the $U(1)$ problem to be solved, we get the standard result:
\beq
\label{chiQCD}
\chi_{QCD} ~ \theta \sim \frac {\theta}{ \sum (2 |H |m_i)^{-1}}  \sim \new{m_{\pi}^2 F_{\pi}^2}  ~\frac{m_u m_d }{(m_u + m_d)^2} ~ \theta
\eeq

In order to make CP violation much smaller than (\ref{chiQCD}) one would need an extremely small $A$, indeed so small that the $U(1)$ problem would remain fully unsolved.

\new{In (\ref{chiQCD}) $F_\pi \simeq 92$ MeV, in the normalization implied by (\ref{Wapp}). For $A \to \chi_{YM}$ the above expression for $\chi_{QCD}$, as well as the results of the next Section, reduce to those known from the large-$N$, $U(N_f)$ chiral Lagrangian \cite{DiVecchia:1980yfw,Witten:1980sp,Leutwyler:1992yt,Kaiser:2000gs,DiVecchia:2017xpu}. What is new here is that they hold beyond the large-$N$ limit, with $\chi_{YM}$ replaced by the quantity $A$ defined in (\ref{Wder2}) (see also \cite{Shore:1991dv,Shore:1999tw}).}

\section{Some amusing $\theta$-dependences}
\label{sec:thetadep}
Although the topic is somewhat outside the main goals of this work we wish to point out some amusing connections between the corrections to the PNGB masses at non-vanishing $\theta$ and the topological susceptibility in QCD.

For the pseudoscalar mesons outside the Cartan subalgebra the only contribution to the mass $\mu_{ij}$ comes from the quadratic term in (\ref{massterm}) and, to leading order in $m$, reads:

\beq
\label{GMORtheta}
\mu_{ij}^2 = 2 F_{\pi}^{-2} |H| \left (m_i \cos \langle \phi_i \rangle +  m_j \cos \langle \phi_j \rangle \right)~;~ i \ne j
\eeq
which extends to finite $\theta$ the famous GMOR relation \cite{Gell-Mann:1968hlm}. At small $\theta$ we can expand the cosines and get:
\begin{eqnarray}
\label{GMORsmtheta}
& \mu_{ij}^2 \sim 2 F_{\pi}^{-2} |H| \left(m_i +  m_j - \frac12 m_i \sin^2 \langle \phi_i \rangle - \frac12  m_j \sin^2 \langle \phi_j \rangle \right)  \nonumber \\
&= 2 F_{\pi}^{-2} |H| (m_i +  m_j) \left(1 - \frac18 \frac{\chi_{QCD}^2}{ (H m_i)(H m_j) } \theta^2 \right) ~;~ i \ne j
\end{eqnarray}
where we have used (\ref{solangles}) and (\ref{Qchi}).

More interesting, perhaps, is the correction to the WV relation. Leaving the generalization to arbitrary masses (where singlet-non singlet mixing occurs) as an exercise, we give here the result for the $\eta'$ mass to order $\theta^2$ in the equal mass limit  
\beq
\label{WVsmtheta}
m_{\eta'}^2 F_{\eta'}^{2} \sim  4 m |H|  + 2 N_f A - \frac{1}{N_f}\chi_{QCD} \theta^2   + \frac{1}{N_f}\frac{1}{A}\left(1- N_f^2 \frac{\new{\hat{Q}^{(3)}(0)}}{A}\right) \chi_{QCD}^2 \theta^2
\eeq
Note that the last contribution is of order $m^2$ and contains \new{the non-universal quantity $\hat{Q}^{(3)}(0)$ which, as in (\ref{Weven}), denotes the third derivative of $\hat{Q}$ with respect to its argument $\theta + Tr \phi$ at fixed $H$, evaluated at zero argument}. On the other hand, since $\chi_{QCD} $ is of order $m$, it can be consistently ignored in the chiral limit.

Our last point concerns the $\theta$ dependence of the $CP$-violating Lagrangian. One finds:
\beq
V_{CPV} = \langle Q \rangle  \left[  Tr \sin \Phi - Tr \Phi \new{\; + \frac{\hat{Q}^{(3)}(0)}{6 A}\,(Tr \Phi )^3} \right] + \dots
\eeq
where the corrections are non universal even at leading order in $\theta$. \new{For the 't Hooft-type actions of Sect.~\ref{sec:tHooft} one has $\hat{Q}^{(3)}(0) = A$ and the bracket reproduces $Tr \sin \Phi - \sin (Tr \Phi)$ to this order, while in the large-$N$ case $\hat{Q}^{(3)}(0)=0$.} 

\section{ A few instructive examples of $\Gamma({\cal M},Q)$}
\label{sec:examples}
In this Section we will show how the general framework outlined in the previous Section gets explicitly realized in three well-known examples:
\begin{itemize}
    \item the conventional 't Hooft large-$N$ limit of QCD;
    \item the original 't Hooft proposal for incorporating (dilute) instanton effects;
    \item the orientifold large-$N$ limit of QCD.
\end{itemize}
In each case we will see their consequences for the $U(1)$ problem and for CP-violating processes.

\subsection{The conventional large-$N$ effective action}
In the large-$N$ limit, quark-gluon mixing as well as flavor mixing and higher order multi-$Q$ correlators are suppressed.
In the general framework discussed in Sect. (\ref{sec: theorem}) this implies:
\begin{itemize}
\item $V_{inv}$ of (\ref{eq:Gammainvesp}) can be restricted to the first two terms:
\beq
\label{eq:VinvN}
   V_{inv.}({\cal M},Q; 0, 0) = V_{inv}({\cal M},0) + \frac{Q^2}{2}  V_{QQ} \equiv V_{inv}({\cal M},0) - \frac{Q^2}{2 \chi_{YM}}
\eeq
where $-(V_{QQ})^{-1}$ is just a constant that gets identified with the topological susceptibility in pure YM theory. While the first term is of ${\cal O}(N)$ (like the mass term) the second is of ${\cal O}(1)$. In general there can be competition between the large-$N$ and chiral limits \cite{DiVecchia:2017xpu};
\item There is a universal kinetic term for all the PNGBs which fixes their common decay constant to be $F_{\pi}$ and of ${\cal O}(\sqrt{N})$.
\end{itemize}

The consequences of these peculiarities of the large-$N$ limit are quite simple.

On the side of the mass spectrum the general result (\ref{Wapp}) becomes:
\beq
 \label{WV}
W = \frac12 \chi_{YM}(Tr \Phi)^2= \frac{\chi_{YM}}{F_\pi^2}
(\sum_{i=1}^{N_f} \Pi_{ii})^2 = \frac{2 N_f }{F_\pi^2}~ \chi_{YM} \frac12 \eta_0^2
\eeq
corresponding to the WV relation \cite{Witten:1979vv,Veneziano:1979ec} i.e. to  a square mass of ${\cal O}(N_f/N)$ in the flavor-singlet channel.

Concerning CP-violation, one finds that the whole series of higher order terms in $Tr \Phi$ written in (\ref{Wodd}) is missing, so that the full CP-odd potential is simply given by:
\beq
\label{CPoddN}
 V_{CPodd} = \langle Q \rangle \times Tr [\sin \Phi - \Phi] =  - \theta ~ \chi_{QCD} \times Tr [\sin \Phi - \Phi]~;~ \Phi = \frac{\sqrt{2}}{F_\pi} \Pi
\eeq
as already known from the literature \cite{DiVecchia:2017xpu}.

\subsection{The 't Hooft effective action}
\label{sec:tHooft}
't Hooft's original effective action for the anomalous piece of $\Gamma$ is:
\begin{equation}
\label{Vt}
V_{an,t} = -\frac{|C|}{2} \left( \det {\cal M} e^{i \theta } + h.c.\right) \ ,
\end{equation}
where the constant $C$, introduced for dimensional reasons, should be proportional to the appropriate power of $\Lambda_{QCD}$. As already mentioned in Sect. (\ref{sec:general}) it looks a priori difficult to obtain such a result from the general expression (\ref{eq:GammaUQan-gen}) after replacing $Q$ with $\hat{Q}$ through its own equations of motion.

In fact one can even prove the following no-go theorem:
 If the invariant part of the potential is of the form
 \beq
 \label{nogo}
 V_{inv} = \tilde{V}_{inv}({\cal M}) + V_{inv} (Q^2) \, ,
 \eeq
i.e. if the $Q$ and ${\cal M}$ dependence is not intertwined,  there is no way to recover 't Hooft's effective Lagrangian after integrating out $Q$. The proof is given in Appendix B.

However, if we take $V_{inv}$  to depend also upon ${\cal M}$ through $|\det {\cal M}|$, a  solution can be found. We need $\partial_Q V_{inv} (Q^2, |\det {\cal M}|) = \arcsin(\frac{Q}{ C|\det {\cal M}|})$, so that:
\beq
\label{Qhatt}
\hat Q = - |C \det {\cal M}| \sin(\theta + \arg \det {\cal M}) = \frac{|C| }{2 i} \left( \det {\cal M} e^{i \theta} - h.c. \right)
\eeq
The solution is:
\begin{equation}
\label{Gt}
V_{inv} (Q^2, |\det {\cal M}|)= - \sqrt{|C \det {\cal M}|^2 - Q^2} - Q \arcsin(\frac{Q}{|C \det {\cal M}|})
\end{equation}

One then verifies that, after integrating out $Q$ (i.e. after setting $Q = \hat{Q}$ in (\ref{Gt})) one obtains 't Hooft's original proposal (\ref{Vt}).

Interestingly, one can define a whole class of generalizations of 't Hooft's  effective action by taking the constant $C$ to be itself a $U(N_f)\otimes U(N_f)$-invariant function of ${\cal M}$. The result is again (\ref{Vt}), but with an extra dependence upon $|{\cal M}|$ and/or $|\det {\cal M}|$. A particularly interesting case is one in which $|C \det {\cal M}|$ is just a constant $A$, so as usual $1/A = -V_{QQ}(0)$, giving
\begin{equation}
\label{Vt1}
V_{gen.t} = -\frac{A}{2} \left(\frac{\det {\cal M}}{|\det {\cal M}|} e^{i \theta } + h.c. \right)
\end{equation}
This example is not in contradiction with the above theorem since the result is not a linear function of $\det {\cal M}$. It is nevertheless interesting that such generalizations of 't Hooft's action exist and satisfy all the constraints.

\noindent 
  {A common characterization of 't Hooft's effective action and its generalizations described  above is that, when seen as functions of $\theta$ and the angles, they are analytic and periodic (with periodicity $2 \pi$) both in $\theta$ and in each $\phi_i$.} This is what distinguishes this class of actions from those of the large-$N$ type which, instead, {are only invariant under a joint shift by $ 2 \pi$ of $\theta$ and one of the angles $\phi_i$. They acquire periodicity in $\theta$ thanks to their multi-branch (logarithmic) structure with branch-change occurring at $\theta = \pi + 2 k\pi$.}
{Conversely, while large-$N$-type effective actions follow from simple, single-valued expressions for $V_{inv}(Q)$ those of 't Hooft's type need a multivalued  $V_{inv}(Q)$. Indeed, if $W$ is  analytic and periodic in its argument $(\theta + Tr \langle \phi \rangle)$, the same is true for its derivative and, by eq. (\ref{Wder}), also $\hat{Q}(\theta + Tr \langle \phi \rangle)$ should be a periodic function. That implies that  $V_{inv,Q}(Q)$ cannot be single-valued.}

 \noindent 
 We will now analyze the implications of this general class of actions for the $U(1) $ and $CP$ problems.
It is straightforward to repeat the reasoning given in Sect.~(\ref{sec: theorem}) to the case of 't Hooft-like actions.
Since the mass term is the same for any model (a great advantage of the 1PI action!), the CP-odd term in (\ref{CPmass}) is unchanged.

Instead, the CP-odd term generated by $W$ is different since it depends on $V_{inv}$. We know, however, that, in full generality, it produces a potential satisfying (\ref{Wder}) with $\hat{Q}$ given by (\ref{Qhatt}). Equation (\ref{Wder})  can be immediately integrated and gives: 
\beq
\label{Woddt}
W({\cal M}) = - |C| |\det {\cal M}| \cos (\theta + Tr \langle \phi \rangle + Tr  \Phi ) = - \langle Q \rangle \sin (Tr \Phi)  + ({\rm CP-even})
\eeq

Consequently, the CP-violating \new{potential} takes the form:
\begin{equation}
\label{CPVt}
\new{V_{CPV}} =  \langle Q \rangle  \times \left(Tr \sin (\Phi) - \sin (Tr \Phi) \right)
\end{equation}

Equation (\ref{Woddt}) shares with the general case the property that it contains no linear term in $\eta'$. However, it has the amusing property that, for $N_f=1$, there is no $CP$ violation in the $\eta'$ sector, something that can be checked directly from (\ref{Vt}). On the other hand, as long as $N_f >1$, CP-odd terms involving only $\eta'$ self-interactions are also present. Clearly (\ref{CPmass})-(\ref{Wodd}) and (\ref{CPVt}) only differ in their predictions about processes involving {\it only} the singlet PNGB. This is understandable since, without taking the large-$N$ limit, one cannot hope to be able to make model-independent predictions for processes involving just that not-particularly-light state\footnote{At large-$N$, instead, soft amplitudes involving $n$-singlets are suppressed by powers of $1/N$ and  get related to higher cumulants of the YM topological charge \cite{Witten:1979vv,Veneziano:1979ec}.}.

On the other hand the reasoning that led us to the identification of $\chi_{QCD}$ in (\ref{Qchi}) still holds as well as the resulting tension between solving the $U(1)$ and $CP$ problems. Indeed one can compute the anomaly induced mass matrix for the (generalized) 't Hooft case and find:
\beq
\label{eta'masst}
V_{mass,t} = |C| |\det {\cal M}|  \cos (\theta + Tr \phi ) \frac12 (Tr \Phi )^2 =  \frac12  |C| |\det {\cal M}| (Tr \Phi )^2 + {\cal O}(\theta^2)
\eeq
Comparing this with (\ref{CPVt}) we see that the coefficient in front of CP violation and the one in front of the $\eta'$ mass term are related by  an extra factor
$\sin (\theta + Tr \langle \phi \rangle )$ in the former. This factor can be evaluated as in (\ref{thetaSigma}) with the same conclusion: The only way to suppress CP-violation is to make the anomaly induced mass matrix much smaller than the one induced by the quark-masses!
\subsection{The orientifold large-$N$ case }
\label{sec:orient}

In the orientifold large-$N$ limit the quark transforms in the two-index (anti)symmetric
representation \cite{Corrigan:1979xf}.  Planar equivalence relates, {at
infinite $N$,  the bosonic subsector of both theories to the one in ${\cal N}=1$ super Yang--Mills
\cite{Armoni:2003fb,Armoni:2003gp,Armoni:2004uu}}; with a small quark mass it corresponds to softly
broken SYM \cite{Masiero:1984ss}.  Since the $U(1)_A$ anomaly does not switch off at large $N$,
this theory provides an analytically controlled realization of the $U(1)$--$CP$ tension.

\new{For one Dirac flavor $q$ in the two-index representation the order parameter is, as in the
rest of the paper, the quark bilinear
\beq
{\cal M}\equiv\bar q_R\,q_L=H\,e^{i\phi},\qquad H<0,
\label{eq:MOrient}
\eeq
with the colour indices summed. The anomalous Ward identity now implies that a rotation of
the phase $\phi$ is equivalent to a shift of $\theta$ by $n_{\cal R}\phi$, so that the anomaly
term of (\ref{eq:GammaUQan-gen}) becomes $\tau\,Q$ with}
\beq
n_{\cal R}\equiv N\mp2,\qquad
\tau=\theta+n_{\cal R}\phi,
\label{eq:tauOrientShort}
\eeq
where the upper sign denotes the antisymmetric representation \new{and} the lower sign the symmetric
\new{one}.  The anomaly-matched orientifold effective
Lagrangian of \cite{Sannino:2003xe} corresponds, in the notation of (\ref{eq:GammaUQan-gen}), to an
invariant potential exactly quadratic in $Q$,
\beq
\begin{split}
V_{inv}({\cal M},Q)&=V_{inv}(|{\cal M}|,0)+\frac{Q^2}{2}V_{QQ}(|{\cal M}|),\\
\new{V_{QQ}}&\new{=-\frac{1}{{\cal A}_{\cal O}(|{\cal M}|)},
\qquad
{\cal A}_{\cal O}(|{\cal M}|)=a_{\cal O}\,({\cal M}^{\dagger}{\cal M})^{2/3}>0,}
\end{split}
\label{eq:VQQOrientShort}
\eeq
\new{where $a_{\cal O}$ is a dimensionless positive constant}\footnote{\new{The effective Lagrangian of
\cite{Sannino:2003xe} is written in terms of the field
${\cal M}_{\cal O}=-\frac{3\lambda}{4N^2}\,{\cal M}$, where
$\lambda\equiv g^2N/(8\pi^2)$ is the 't~Hooft coupling, and of a dimensionless coupling
$\alpha$; this relation follows from matching the mass term to (\ref{eq:GammaUQan-gen}).
The constant $\Lambda$ is the dynamically generated scale of the theory, normalized such that
$\langle{\cal M}_{\cal O}\rangle=\Lambda^3$ in the massless theory, so that
$|H|=\frac{4N^2\Lambda^3}{3\lambda}$. Furthermore
$a_{\cal O}=\frac{8\alpha}{9}\frac{f(N)}{N^2}\left(\frac{3\lambda}{4N^2}\right)^{4/3}$ and
$F_{\eta_0}^2=4f(N)\Lambda^2/\alpha$, with $f(N)\to N^2$. The singlet mass (\ref{eq:metaOrient}) is
then $m_{\eta_0}^2=\frac{4\alpha^2}{9}\frac{n_{\cal R}^{\,2}}{N^2}\Lambda^2$.}}.
Eliminating $Q$ gives
\beq
\new{\hat Q=\frac{\tau}{V_{QQ}}=-{\cal A}_{\cal O}\,\tau,
\qquad
W=-\frac{\tau^2}{2V_{QQ}}=\frac{{\cal A}_{\cal O}}{2}\tau^2,}
\label{eq:WOrientShort}
\eeq
so that ${\cal A}_{\cal O}$ is the orientifold value of the coefficient $A$ of (\ref{Wapp}).
Thus the anomaly contribution $W$ is purely quadratic and generates no higher CP-odd singlet tower
of its own.  This does not mean that the full orientifold theory has no higher CP-odd singlet
interactions: after vacuum alignment they are generated by the mass term, as shown below.  The
dependence of $V_{QQ}$ on $|{\cal M}|$ intertwines the radial and $Q$ sectors, precisely the
structure that evades the no-go theorem of Appendix~B once $|{\cal M}|$ is frozen in the
non-linear realization.  More explicitly, before freezing the radial mode one should regard
${\cal A}_{\cal O}={\cal A}_{\cal O}(|{\cal M}|)$; it becomes the constant
\new{${\cal A}_{\cal O}(|H|)$} only after radial minimization.

\new{At large $N$ the condensate and the singlet decay constant, defined as in (\ref{Wapp})
through $\delta\phi=\sqrt2\,\eta_0/F_{\eta_0}$ for the fluctuation $\delta\phi$ of the phase,
scale as $|H|={\cal O}(N^2)$ and $F_{\eta_0}^2={\cal O}(N^2)$, while
${\cal A}_{\cal O}(|H|)={\cal O}(N^0)$. The anomaly-induced singlet mass, the analog of
(\ref{Wapp}),
\beq
m_{\eta_0}^2=\frac{2{\cal A}_{\cal O}\,n_{\cal R}^{\,2}}{F_{\eta_0}^2}\,,
\label{eq:metaOrient}
\eeq
is therefore} ${\cal O}(N^0)$: the large-$N$ suppression of the conventional 't~Hooft limit is absent.

\new{With $H$ frozen at its vacuum value, the mass term (\ref{massterm}) reads
\beq
V_m=2Hm\cos\phi=-c_m\cos\phi,\qquad
c_m=2|H|m\,.
\label{eq:massOrientShort}
\eeq}

We now make the local, small-$\theta$ vacuum-alignment argument explicit.  Let
\beq
V(\phi;\theta)=\frac{{\cal A}_{\cal O}}{2}\tau^2-c_m\cos\phi.
\label{eq:phasePotentialOrient}
\eeq
Writing the aligned vacuum as $\langle\phi\rangle=\phi_0$, and setting
$\tau_0\equiv\theta+n_{\cal R}\phi_0$, stationarity gives
\beq
{\cal A}_{\cal O}n_{\cal R}\tau_0+c_m\sin\phi_0=0,\qquad
\langle Q\rangle=-{\cal A}_{\cal O}\tau_0,\qquad
c_m\sin\phi_0=n_{\cal R}\langle Q\rangle .
\label{eq:alignmentOrient}
\eeq
At small $\theta$,
\beq
\langle Q\rangle=-\new{\chi_{\cal O}}\,\theta+{\cal O}(\theta^3),
\qquad
\new{\chi_{\cal O}}
=\frac{{\cal A}_{\cal O}c_m}{c_m+{\cal A}_{\cal O}n_{\cal R}^{\,2}}
=\left[\new{\frac{1}{{\cal A}_{\cal O}}}+\frac{n_{\cal R}^{\,2}}{c_m}\right]^{-1},
\label{eq:chiOrientShort}
\eeq
\new{where $\chi_{\cal O}$ is the topological susceptibility of the full orientifold theory, the
analog of $\chi_{QCD}$ of (\ref{Qchi}). It} is the total $\theta$-derivative along the aligned vacuum, as in (\ref{Qchi}) and
(\ref{thetaSigma}); it is distinct from the fixed-angle curvature
${\cal A}_{\cal O}n_{\cal R}^{\,2}$ that sets the anomaly-induced singlet mass.  For a recent
effective-field-theory review of the strong-$CP$ problem, topological susceptibility and the QCD
axion, see Ref.~\cite{Sannino:2026wgx}.  In the small-mass limit,
\beq
\new{\chi_{\cal O}}
=\frac{c_m}{n_{\cal R}^{\,2}}
\left[1+{\cal O}\!\left(\frac{c_m}{{\cal A}_{\cal O}n_{\cal R}^{\,2}}\right)\right]
=\new{\frac{2|H|m}{N^2}}
\left[1+{\cal O}\!\left(N^{-1}\right)\right]+{\cal O}(m^2).
\label{eq:chiScalingOrientShort}
\eeq
Thus the $N^2$ \new{of the condensate} cancels $n_{\cal R}^{\,2}=(N\mp2)^2$:
\new{$\chi_{\cal O}$}$={\cal O}(mN^0)$ in both representations and vanishes in the chiral limit.

Shifting about the aligned vacuum,
\beq
\phi=\phi_0+\delta\phi,\qquad
\tau=\tau_0+n_{\cal R}\delta\phi,
\label{eq:shiftOrient}
\eeq
the odd part of the mass potential is
\beq
V_{m,\mathrm{odd}}
=c_m\sin\phi_0
\left[\delta\phi-\frac{(\delta\phi)^3}{3!}
+\frac{(\delta\phi)^5}{5!}-\cdots\right].
\label{eq:massOddOrient}
\eeq
The quadratic anomaly potential supplies only the linear odd term
${\cal A}_{\cal O}n_{\cal R}\tau_0\delta\phi$.  Its sum with the first term in
(\ref{eq:massOddOrient}) vanishes by (\ref{eq:alignmentOrient}).  Therefore the tadpole cancels,
while the cubic and higher odd terms survive:
\beq
V_{CP\mathrm{-odd}}
=n_{\cal R}\langle Q\rangle
\left[-\frac{(\delta\phi)^3}{3!}
+\frac{(\delta\phi)^5}{5!}-\cdots\right].
\label{eq:CPoddOrient}
\eeq
\new{In terms of the canonically normalized fluctuation
$\eta_0=F_{\eta_0}\delta\phi/\sqrt{2}$, the first surviving term of the $CP$-odd potential is
\beq
V_{CP}^{(3)}
=\frac{\sqrt{2}\,n_{\cal R}\,\chi_{\cal O}}{3F_{\eta_0}^3}\,
\theta\,\eta_0^3+{\cal O}(\theta^3),
\label{eq:cubicCPOrient}
\eeq}
which is ${\cal O}(m)$ and vanishes in the chiral limit.  This is the qualitative difference from
one-flavor QCD described by the full 't~Hooft cosine.  There the anomaly and mass terms have the
same phase frequency, so their complete odd sine series is proportional to the alignment equation
and all odd \new{$\eta_0$} self-interactions cancel.  In the orientifold theory with a quadratic anomaly
potential, only the tadpole cancels, as in the large $N$ example. 

\section{Summary and Discussion}
\label{sec:discussion}

In this paper we have tackled two important aspects of non-perturbative QCD, the $U(1)$ and Strong-$CP$ problems, as well as their relationship, using the powerful tools of one-particle-irreducible ($1$PI) effective actions $\Gamma_{1PI}$ defined as functionals of a convenient set of gauge invariant operators $O_i$ via a standard Legendre transformation. The approach is in principle exact, but its success depends on asking the right questions, on making a judicious choice of the operators $O_i$, and on being able to justify certain assumptions on that part of $\Gamma$ which is not fixed by general arguments.

In the case at hand we have identified the set of convenient operators as consisting of the scalar and pseudoscalar quark bilinear ${\cal M}_{ij} \equiv \bar{q}_{R,i}q_{L,j} $ (with $i, j = 1, 2 \dots N_f$) and its hermitian conjugate ${\cal M}^{\dagger}$, and of the topological charge density $Q \equiv \frac{1}{32\pi^2} G \tilde{G}$. This choice completely fixes the dependence of $\Gamma$ on all the QCD parameters, the quark mass matrix $m$ and the vacuum  angle, $\theta$, leaving outside only the overall RGI scale $\Lambda_{QCD}$.

Using the known symmetries of QCD in the massless limit, as well as the exact form of the $U(1)$ axial anomaly, fixes $\Gamma$ up to a $CP$-even and $U(N_f)_R \otimes U(N_f)_L$-invariant term $\Gamma_{inv.}({\cal M},Q)$ (see eq. (\ref{eq:GammaUQan})). Furthermore, one can directly show that, up to inconsequential local field redefinitions, the mass matrix can be taken to be real and diagonal through the usual replacement of $\theta$ by $\bar{\theta} \equiv \theta - \arg \det m$.

Although, as shown in a few examples, some detailed predictions do depend on the explicit form of $\Gamma_{inv.}( {\cal M},Q)$, we have  shown that, in the low-energy, small-quark-masses and small
$\bar{\theta}$ limit, all relevant predictions are robust wrt changes of $\Gamma_{inv.}$ and depend essentially only upon a single quantity $A$ defined as:
\beq
\label{Adef}
A \equiv - \left(\frac{\partial^2 V_{inv}}{(\partial Q)^2 } \right)^{-1}_{(Q=0)}
\eeq
 where $- V_{inv.}( {\cal M},Q)$, is the zero momentum limit of  $\Gamma_{inv.}( {\cal M},Q)$.

 $A$ is some sort of ``quenched" topological susceptibility $\chi_{t,q}$ and indeed one can prove that, in 't Hooft's large-$N$ limit, it is just the topological susceptibility of Yang-Mills theory. In any case a non vanishing (and positive) value of $A$ implies a contribution to the flavor-singlet entry of the pseudo-Nambu-Goldstone-boson (PNGB) mass matrix potentially solving the $U(1)$ problem a la Witten-Veneziano \cite{Witten:1979vv,Veneziano:1979ec}. The tension mentioned in the title of the paper is just the statement that, in order to solve the strong-CP problem, one would need $A$ to be extremely small, so small that its contribution to the PNGB mass matrix would be negligible compared to the one due to the quark masses, therefore leaving the $U(1)$ problem unsolved.

 The above statement is made quantitatively precise by showing that the strength of $CP$ violation depends on the full unquenched topological susceptibility of QCD, $\chi_{t,QCD}$ which is suppressed wrt $A$ by powers of the light-quark masses and actually becomes independent of $A$ when $A$ is large enough to solve the $U(1)$ problem. In other words one falls back on the conclusion that, if the $U(1)$ problem is solved, the CP problem can only be avoided, within QCD, by having at least one massless quark.

 This is the most important --albeit largely expected-- conclusion of this work. We also show how things work out in more detail for various choices of $V_{inv.}$ like the conventional  large-$N$ case, the orientifold large-$N$ case or the original 't Hooft instanton effective action. The detailed results differ, particularly when the flavor-singlet would-be NGB is concerned, but the above conclusion on the $U(1)$-CP tension remains unchanged.

As a side remark, we also show that the small-$\theta$ correction to the PNGB masses and those to $CP$-violating amplitudes are all controlled by the size of $\chi_{t,QCD}$.

In this paper we have mainly focused on the physically relevant case of small $\bar\theta$, and have not considered
the vacuum structure for larger $\bar\theta$ where it is known that many interesting features emerge,
especially for $\bar\theta = \pi$ where CP can be spontaneously broken \cite{Dashen:1970et}.    As a further result of our analysis we have shown that two approaches to the $U(1)$ and CP problems, i.e. the one in~\cite{Christos:1984tu} based on the large $N$ expansion and the one in~\cite{tHooft:1986ooh} based on instantons, originally thought to be incompatible, can both emerge from the 1PI approach. 

For small $\bar\theta$, the vacuum equations (\ref{solangles}) are essentially identical for the large-$N$ and 't Hooft 
models. This is not guaranteed for larger values of $\bar\theta$ and $\sum_i \langle \phi_i \rangle$. However, their near cancellation in the small mass limit in eq.~(\ref{thetaSigma}) allows one to extend many results away from the small $\bar{\theta}$ limit. We plan to return to the issue of the vacuum structure and the spectrum of the theory in a forthcoming paper \cite{DiVecchia:2026QCD}.

\new{Finally, it is useful to state where the claims of
\cite{Ai:2020ptm,Nakamura:2021meh,Williams:2026tuw} depart from the framework used here.
Our results are conditional: with $\bar\theta$ a parameter of the action, (i) the lemma of
Sect.~\ref{sec:lemma} implies that $\theta$ enters $\Gamma$ only as a constant source for
$Q$, and (ii) $\theta$ is RG-invariant. Together they give
$\langle Q\rangle=-\chi_{QCD}\,\bar\theta$, where, at non-zero quark masses, $\chi_{QCD}$
vanishes only if $A$ does. Hence a scenario in which $CP$ is conserved at
$\bar\theta\neq0$ while the $\eta'$ keeps its anomalous mass has to give up one of the two
premises. In \cite{Ai:2020ptm} the infinite-volume limit is taken before the sum over
topological sectors, which amounts to a definition of the functional integral in which
$\theta$ is no longer the source conjugate to $Q$, so that (i) does not apply. In
\cite{Nakamura:2021meh} the renormalized $\theta$ flows to zero in the infrared, in
contrast with (ii). Within our framework the only screening of $\theta$ is the partial one
by the quark phases, (\ref{thetaSigma}), which becomes complete only for a massless quark.
We do not discuss here the merits of these proposals. Our analysis does not exclude
$\bar\theta=0$, whether because no such parameter needs to be introduced
\cite{Williams:2026tuw} or because of additional dynamics, and all our results on the
$U(1)$ problem hold in that case; it shows that, once $\bar\theta\neq0$ is allowed, $CP$
violation of strength $\chi_{QCD}\,\bar\theta$ necessarily follows whenever the $U(1)$
problem is solved.}

{Many of the questions addressed  in this paper can also be  investigated in a  two-dimensional $CP^{N-1}$ model with fermions where, unlike in QCD, explicit calculations can be performed \cite{DiVecchia:2026cpn}.  They confirm, mutatis mutandis,
the conclusions we have reached here on the tension between the $U(1)$ and strong-CP problems.}

\section*{Acknowledgments}
The work of F.S.\ is partially supported by the Carlsberg Foundation,
Semper Ardens grant CF22-0922. G.M.S. is grateful to TH Division, CERN for hospitality as a Visiting Scientist. We also wish to acknowledge an interesting exchange with Ken Konishi on the (non)renormalization of $\theta$. \new{We thank the referee for valuable comments that helped us improve the paper.}

\clearpage
\setcounter{section}{0}
\renewcommand{\thesection}{\Alph{section}}
\renewcommand{\theequation}{\thesection\arabic{equation}}
\makeatletter
\@addtoreset{equation}{section}
\@ifundefined{theHsection}{}{\renewcommand{\theHsection}{appendix.\thesection}}
\makeatother
\section*{Appendix}

\section{Vacuum alignment and linear vs.\ non-linear realization}

Although these results are likely known in the literature, we include a short proof for completeness.

Concerning vacuum alignment the proof is simple. For a real diagonal mass matrix the explicit symmetry-breaking potential is:
\beq
\label{masspot}
V_{mass} = 2 \sum_i m_i {\cal M}_{ii}
\eeq
 {Likewise, the anomaly potential depends only on the singlet combination $Tr \Phi$ (or equivalently on $\det\Phi)$ and is therefore insensitive to the off-diagonal matrix elements.}
As a consequence the variation of the potential with respect to the off-diagonal entries of ${\cal M}$ receives no contribution from either the mass or the anomaly terms and can only come from varying $V_{inv}$.  The invariant potential cannot contain terms linear in the off-diagonal components of $\Phi$. Every invariant is built from traces and determinants, and therefore every dependence on an off-diagonal component starts at least quadratically.

Therefore $\langle {\cal M}_{ij} \rangle = \delta_{ij} {\cal M}_{ii}$ is a possible solution although, as mentioned in the text, not necessarily the only one. Since the polar decomposition
${\cal M}=HU$ is unique for non-singular ${\cal M}$, the diagonal nature  of $\langle {\cal M} \rangle$ immediately
implies that both $\langle H \rangle$ and $\langle U \rangle$ are diagonal.  

The claimed equivalence of the linear vs.\ non-linear realization is less obvious. It has to do with the expectation value of the Hermitian matrix $H$ in the presence of a small mass perturbation. The anomaly term does not play any role here since  $V_{an}$ depends only on $U$. On the other hand the mass term is linear in the radial fields, $V_m \sim \sum_i (m_i H_{ii} \cos \phi_i)$. Since the radial modes are massive, the minimum is shifted by  $\delta \langle H_{ii} \rangle ={\cal O}(m_i)$, while the heavy masses remain of order $\Lambda_{QCD}$.

By the assumption of SSB down to $U_V(N_f)$ in the massless limit, $V_{inv}$ has a symmetry breaking minimum proportional to the unit matrix:
$\langle H_{ij} \rangle = \delta_{ij} H$. As a result of the mass perturbation this minimum will be shifted by ${\cal O}(\frac{m_i}{\Lambda_{QCD}})$ since, by assumption, the $H_{ii}$ directions are not flat. {Evaluating the effective potential on the shifted minimum,
$V_{eff}= V_{inv} (H_{cl}) + V_m(H_{cl})$
the terms linear in $\delta H$ cancel identically because $H_{cl}$ satisfies the stationarity equations,
$\partial V /\partial H_{ii} =0$.
Consequently, the first non-trivial correction generated by integrating out the radial modes is quadratic in the quark masses.}
In other words, the leading order-in-$m$ predictions are the same whether we use the linear or the non-linear parametrization for SSB. {Therefore integrating out the heavy radial modes merely generates higher-order symmetry-breaking operators of order $m^2, m^3, \dots$, while the leading ${\cal O}(m) $ chiral potential is identical to that obtained in the nonlinear realization.} {Diagrammatically, the cancellation of the terms linear in $\delta H$ is equivalent to the cancellation of the tadpoles of the heavy radial fields. The remaining effects of integrating out $H$ are local higher-dimensional operators suppressed by the heavy scalar masses.}

\section{No-go theorem for 't Hooft's anomaly term and its evasion}
Proof: Define the complex variable $z \equiv {\det} {\cal M} e^{i \theta}$. For 't Hooft's action $Q$ must be proportional to $ Im\,z$. However, {if we start with the ansatz (\ref{nogo}),} the equation of motion for $Q$ would  read:
\begin{equation}
\partial_Q V_{inv} (\hat{Q}^2) = \frac{1}{2 i} \log (z/z^*) = \arg z \,.
\end{equation}
{Such an equation forces $\hat Q$ to depend on $z$ only through its phase, whereas 't Hooft's action requires $\hat Q \propto {\rm Im}\, z = |z| \sin (\arg z)$, which depends also on $|z|$: consequently, no strict 't Hooft solution exists if we start from (\ref{nogo}). As it is the case with most no-go theorems also in this case it is possible to evade it as shown in the main text. This can be achieved in two ways: (i) By starting from a $V_{inv}$ potential having an entangled dependence upon $Q$ and $|\det {\cal M}|$ in order to recover exactly 't Hooft's effective action; (ii) Sticking to a $V_{inv}$ which is of the form (\ref{nogo}) to generate an effective action that generalizes 't Hooft's in a rather inessential/mild way.

\bibliographystyle{apsrev4-2}
\bibliography{U_1_-CP-Final}

\end{document}